\documentclass[conference,a4paper]{IEEEtran}
\usepackage{xcolor}
\usepackage{balance}

\ifCLASSINFOpdf
  \usepackage[pdftex]{graphicx}
\else
  \usepackage[dvips]{graphicx}
\fi
\ifCLASSOPTIONcompsoc
 \usepackage[caption=false,font=normalsize,labelfont=sf,textfont=sf]{subfig}
\else
 \usepackage[caption=false,font=footnotesize]{subfig}
\fi
\usepackage{ucs}
\usepackage{amsmath}
\usepackage{amsfonts}
\usepackage{amssymb}
\usepackage{makeidx}
\usepackage{ulem} 
\usepackage{bm} 
\usepackage{dsfont}
\usepackage{tipa} 
\usepackage{array}
\usepackage{bigints}
\usepackage{enumitem} 
\usepackage{environ} 
\usepackage{longtable}

\allowdisplaybreaks 

	\renewcommand{\emph}[1]{\textit{#1}}
	\newcommand{\ie}{\textit{i.e.} }
	\newcommand{\eg}{\textit{e.g.} }

	\newenvironment{soe}
				{\left\lbrace\begin{array}{@{}l@{}}}
				{\end{array}\right.}
	\NewEnviron{subs}[1]
				{
				\begin{subequations}
				\label{#1}
				\begin{align}
				\BODY
				\end{align}
				\end{subequations}
				}
	\newcommand{\vect}[1]{\boldsymbol{\mathbf{#1}}} 
	
	\newcommand{\dir}[1]{\hat{\boldsymbol{\mathbf{#1}}}} 
	\newcommand{\mat}[1]{{\underline{\underline{#1}}\,}}
	
	\newcommand{\rpos}{\mathbf{r}}

\def\XXint#1#2#3{{\setbox0=\hbox{$#1{#2#3}{\int}$}
     \vcenter{\hbox{$#2#3$}}\kern-.5\wd0}}

\newcommand{\xdir}{$\dir{x}$ }
\newcommand{\ydir}{$\dir{y}$ }
\newcommand{\zdir}{$\dir{z}$ }

	\newcommand{\Ptot}{P_\text{tot}}
	
\newcommand{\GF}{\underline{\underline{\Gamma}}}
	\newcommand{\GFej}{{\underline{\underline{\Gamma}}^{EJ}}}
	
	\newcommand{\GFhj}{{\underline{\underline{\Gamma}}^{HJ}}}

\newcommand{\rone}{\rpos_1}
\newcommand{\rtwo}{\rpos_2}
\newcommand{\rthree}{\rpos_3}
\newcommand{\rfour}{\rpos_4}

\newcommand{\fieldcorr}{\mat{C}_\text{em}(\rone, \rtwo)}
	\newcommand{\fieldcorrEE}{\mat{C}_\text{em}^{EE}(\rone, \rtwo)}
	\newcommand{\fieldcorrEH}{\mat{C}_\text{em}^{EH}(\rone, \rtwo)}
	\newcommand{\fieldcorrHE}{\mat{C}_\text{em}^{HE}(\rone, \rtwo)}
	\newcommand{\fieldcorrHH}{\mat{C}_\text{em}^{HH}(\rone, \rtwo)}
	\newcommand{\fieldcorrij}{\mat{C}_\text{em}^{ij}(\rone, \rtwo)}
	\newcommand{\fieldcorrnor}{\mat{C}_\text{em}}

\newcommand{\Cabs}{\mat{C}_\text{abs}(\rone, \rtwo)}
\newcommand{\Cabsnor}{\mat{C}_\text{abs}}
	\newcommand{\Cabsjj}{\mat{C}_\text{abs}^{JJ}(\rone, \rtwo)}
	\newcommand{\Cabsjm}{\mat{C}_\text{abs}^{JM}(\rone, \rtwo)}
	\newcommand{\Cabsmj}{\mat{C}_\text{abs}^{MJ}(\rone, \rtwo)}
	\newcommand{\Cabsmm}{\mat{C}_\text{abs}^{MM}(\rone, \rtwo)}
	
		\newcommand{\Cabsjjconj}{\big(\mat{C}_\text{abs}^{JJ}\big)^*(\rone, \rtwo)}
		\newcommand{\Cabsjmconj}{\big(\mat{C}_\text{abs}^{JM}\big)^*(\rone, \rtwo)}
		\newcommand{\Cabsmjconj}{\big(\mat{C}_\text{abs}^{MJ}\big)^*(\rone, \rtwo)}
		\newcommand{\Cabsmmconj}{\big(\mat{C}_\text{abs}^{MM}\big)^*(\rone, \rtwo)}

\renewcommand{\matrix}[1]{\left[\begin{array}{@{}l@{}} #1 \end{array}\right]}

\renewcommand{\iiint}{\int \hspace{-0.25cm} \int \hspace{-0.25cm} \int}

\newcommand{\Imat}{\mat{1}}

\newcommand{\Js}{\vect{J}_s}

\renewcommand{\Im}{\text{Im}}

\newcommand{\Zij}[1]{\mat{Z}_{#1}}
	\newcommand{\Zijin}[1]{{\Zij{#1}^\text{in}}}
	\newcommand{\Zijout}[1]{{\Zij{#1}^\text{out}}}
	
\renewcommand{\eqref}[1]{(\ref{#1})}

\begin{document}
%
\title{Modal characterization of thermal emitters using the Method of Moments}

\author{\IEEEauthorblockN{
Denis Tihon\IEEEauthorrefmark{1},   
Stafford Withington\IEEEauthorrefmark{1},   
Christophe Craeye\IEEEauthorrefmark{2}
}                                     
\IEEEauthorblockA{\IEEEauthorrefmark{1}
Cavendish Laboratory, University of Cambridge, Cambridge, UK.}
\IEEEauthorblockA{\IEEEauthorrefmark{2}
ICTEAM Institute, Universit\'{e} catholique de Louvain, Louvain-la-Neuve, Belgium.}
 \IEEEauthorblockA{ \emph{*corresponding author: dt501@cam.ac.uk} }
}



\maketitle

\begin{abstract}
Electromagnetic sources relying on spontaneous emission are difficult to characterize without a proper framework due to the partial spatial coherence of the emitted fields. In this paper, we propose to characterize emitters of any shape through their natural emitting modes, \ie a set of coherent modes that add up incoherently. The resulting framework is very intuitive since any emitter is regarded as a multimode antenna with zero correlation between modes. Moreover, for any finite emitter, the modes form a compact set that can be truncated. Each significant mode corresponds to one independent degree of freedom through which the emitter radiates power. The proposed formalism is implemented using the Method of Moments (MoM) and applied to a lossy sphere and a lossy ellipsoid. It is shown that electrically small structures can be characterized with a small number of modes, and that this number grows as the structure becomes electrically large.
\end{abstract}

\vskip0.5\baselineskip

%

\section{Introduction}
Electromagnetic sources at infrared and optical wavelengths rely either on stimulated or spontaneous emissions. Stimulated emission can only be obtained under stringent conditions. However, the resulting fields are highly coherent and can be easily manipulated. Spontaneous emission is much easier to implement, but it produces partially coherent fields which are difficult to handle. 

However, there is a plethora of applications that would benefit from an increased control over partially coherent fields, such as radiative cooling \cite{21}, thermophotovoltaic devices \cite{22} or the design of cheap but well controlled sources for photonic circuits, Optical Wireless Communication (OWC) \cite{1}, inter and intra-chip communication using photonics \cite{2}, satellite optical communication \cite{3} or LED beamforming \cite{4,5}.

Sources based on spontaneous emission emit fields due to thermally driven random fluctuations that appear in their volume. In order to deal with these stochastic fields, one has to use a second-order correlation function that describes the coupling between the fields at different locations \cite{9}. However, the use of a correlation function to describe the fields hinders the straightforward use of antenna theory.

{ 
A technique has been proposed in \cite{6,7} to efficiently compute the near-field radiative heat transfer between bodies. This technique is combined with surface and volume integral equation solvers to provide the power radiatively transferred between bodies of complex geometries. However, the information about the spatial correlation of the fields is lost in the procedure.
}

{ 
This spatial correlation function can retrieved using reciprocity. It can be shown that the intensity of the fields emitted in one location are related to the total power absorbed by the structure when it is illuminated by a source at the same location \cite{10}. This reasoning can be extended to a pair of sources to obtain the full correlation function of the fields \cite{Rytov}. However, manipulating such a field correlation function may be complicated. In \cite{16}, it is proposed to decompose the partially coherent fields into an incoherent superposition of fully coherent modes. 
}

{ 
In this paper, we propose a framework that combines these techniques, where a thermal emitter is characterized using a finite set of emitting modes. Each mode corresponds to a fully coherent field distribution. The total fields emitted are made of an incoherent superposition of these modes. In this way, the emitter can be treated as a classical multi-mode antenna. The emissive modes are retrieved from the response of the structure when it is illuminated by external sources thanks to reciprocity \cite{Rytov}.
}

In Section 2, we derive the correlation function of the thermal fields emitted by a lossy structure and how the emissive modes of the structure can be found using Wolf's modal decomposition. Then, in Section 3, we show how this correlation function and emissive modes can be computed using the MoM. Last, the proposed method is tested on a lossy sphere and a lossy ellipsoid, validating the formalism used. 

\section{Computation of thermally emitted fields}
Consider a lossy volume at constant uniform temperature $T$ located in free-space. The volume is characterized by real constant permittivity, permeability and conductivity $\varepsilon$, $\mu$ and $\sigma$, respectively. The problem studied has been kept simple for clarity, limiting ourselves to ohmic loss only. Due to its finite temperature and the presence of losses, and therefore damping, random fluctuations of currents $\Js(\rpos)$ will appear in the volume and radiate fields. The goal of this Section is to characterize the correlation in these fields averaged over time. Since the proposed method is a spectral method, an $\exp(j\omega t)$ time dependency of the fields and currents is assumed. The frequency dependence of the different quantities hereafter will be implicit. We consider that the average is computed over a sufficient amount of time, with a sufficient duration between consecutive measurements, so that the field correlation between different frequencies vanishes and the time average can be estimated using an ensemble average (ergodicity assumption) \cite{9}.

We first consider the dyadic correlation function of the fields $\fieldcorr$ given by 
\begin{align}
\label{eq:02}
\fieldcorr 
&= 
\left\langle 
	\matrix{\vect{E}\\ \vect{H}}(\rone) 
	\cdot 
	\matrix{\vect{E} \\ \vect{H}}^\dagger(\rtwo)
	\right\rangle, 
\\
&\triangleq
\matrix{\fieldcorrEE \hspace{0.3cm} \fieldcorrEH \\ ~\\ \fieldcorrHE \hspace{0.3cm} \fieldcorrHH}.
\end{align}
with $\vect{E}$ and $\vect{H}$ the electric and magnetic fields, the superscript $\dagger$ corresponding to the transpose conjugate operation and the angle bracket $\langle a \rangle$ denoting the ensemble average of $a$. 

These electric and magnetic fields are generated by the random fluctuations inside the lossy volume, which can be highlighted using the dyadic Green's functions \cite{26}:
\begin{equation}
\label{eq:03}
\begin{soe}
\vect{E}(\rone) = \GF^{EJ}(\rone, \rtwo) \cdot \vect{J}(\rtwo), \\
\vect{H}(\rone) = \GF^{HJ}(\rone, \rtwo) \cdot \vect{J}(\rtwo).
\end{soe}
\end{equation}
with $\GFej(\rone, \rtwo)$ and $\GFhj(\rone, \rtwo)$ describing the electric and magnetic fields generated in $\rone$ by electric currents located in $\rtwo$.

In the rest of the development, we will focus our attention to the $\fieldcorrEE$ term. However, the other terms can be treated in a similar way.
Substituting \eqref{eq:03} into \eqref{eq:02} and integrating the contributions of all the random sources in the volume gives
\begin{equation}
\begin{split}
\fieldcorrEE &=
\Bigg\langle 
	\Big[ \iiint_\Omega \GFej(\rone, \rthree) \cdot \Js(\rthree) dV(\rthree) \Big] \\
& \cdot \Big[\iiint_\Omega \GFej(\rtwo, \rfour) \cdot \Js(\rfour) dV(\rfour) \Big]^\dagger
\Bigg\rangle,
\end{split}
\end{equation}
with $\Omega$ the volume occupied by the emitter.

The convolution of a current distribution by the Green's function is a linear operation, so that the ensemble average and convolution operations order can be swapped:
\begin{equation}
\label{eq:04}
\begin{split}
\fieldcorrEE & = \iiint_\Omega \iiint_\Omega
\GFej(\rone, \rthree) 
\\
& \hspace{-1.5cm}
\cdot 
\Big\langle \vect{J}(\rthree) \cdot \vect{J}^\dagger(\rfour) \Big\rangle
\cdot 
\big(\GFej\big)^\dagger(\rtwo, \rfour)
~ dV(\rthree) ~ dV(\rfour).
\end{split}
\end{equation}

The great advantage of \eqref{eq:04} is that the correlation of the fields is expressed as a function of the correlation of the fluctuating currents \cite{26}, which generally rely on local phenomena and can be expressed using the fluctuation-dissipation theorem as \cite{11,10}
\begin{equation}
\label{eq:05}
\Big\langle
\vect{J}(\rthree) \cdot \vect{J}^\dagger(\rfour) 
\Big\rangle
=
2 \sigma \Theta(T) \delta(\rthree-\rfour) \Imat,
\end{equation}
with $\Imat$ the 3x3 identity matrix, $\delta$ the Dirac delta function and $\Theta(T)$ that is given by
\begin{equation}
\Theta(T) = \dfrac{\hbar \omega}{\exp\bigg(\dfrac{\hbar \omega}{k_B T}\bigg) -1}.
\end{equation}
$k_B$ is the Boltzmann constant and $\hbar$ the reduced Planck constant.

Substituting \eqref{eq:05} into \eqref{eq:04} and simplifying the integrals gives
\begin{equation}
\begin{split}
\fieldcorrEE & = 
2 \sigma \Theta(T) 
\\ & \hspace{-0.5cm} \times
 \iiint_\Omega
\GFej(\rone, \rthree) \cdot \big(\GFej\big)^\dagger(\rtwo, \rthree)
dV(\rthree).
\end{split}
\end{equation}

Similarly, one can develop the other blocks of \eqref{eq:02} to obtain
\begin{equation}
\label{eq:06}
\begin{split}
\fieldcorrij & = 
2 \sigma \Theta(T) 
\\ & \hspace{-0.5cm} \times 
\iiint_\Omega
\GF^{iJ}(\rone, \rthree) \cdot \big(\GF^{jJ}\big)^\dagger(\rtwo, \rthree)
dV(\rthree).
\end{split}
\end{equation}

To summarize, entry $(k,l)$ of $\fieldcorrij$ in \eqref{eq:06} corresponds to the correlation between the $k$-component of the fields of type $i$ in position $\rone$ and the $l$-component of the fields of type $j$ in position $\rtwo$, $i$ and $j$ corresponding to electric and/or magnetic fields and $k$ and $l$ corresponding to any of the three directions of space (\eg \xdir, \ydir and \zdir directions using a Cartesian coordinate system).

Exploiting Lorentz reciprocity, it can be shown that the integral in \eqref{eq:06} can be evaluated considering the power absorbed by the structure when it is excited by impressed electric and/or magnetic currents in positions $\rone$ and $\rtwo$. 

We define the spatial correlation function $\Cabs$ such that the total power $\Ptot$ absorbed by the structure when illuminated by any electric and magnetic currents $[\vect{J}_1^T, \vect{M}_1^T]^T$ and $[\vect{J}_2^T, \vect{M}_2^T]^T$ in positions $\rone$ and $\rtwo$, respectively, reads \cite{15}
\begin{equation}
\Ptot = 
\sum_{i,j = 1}^{2}
\matrix{\vect{J}_i \\ \vect{M}_i}^\dagger
\cdot 
\Cabsnor(\rpos_i, \rpos_j)
\cdot
\matrix{\vect{J}_j \\ \vect{M}_j},
\end{equation}
with $\mat{A}^T$ the transpose of $\mat{A}$. Splitting the $\Cabs$ function into 4 blocks
\begin{equation}
\label{eq:12}
\Cabs 
=
\matrix{\Cabsjj \hspace{0.3cm} \Cabsjm \\ ~ \\ \Cabsmj \hspace{0.3cm} \Cabsmm},
\end{equation}
it can be shown that \cite{Rytov}
\begin{equation}
\label{eq:10}
\begin{soe}
\fieldcorrEE = 4 \Theta(T) \Cabsjjconj \\ ~ \vspace{-0.2cm} \\
\fieldcorrEH = - 4 \Theta(T) \Cabsjmconj \\ ~ \vspace{-0.2cm} \\
\fieldcorrHE = - 4 \Theta(T) \Cabsmjconj \\ ~ \vspace{-0.2cm} \\
\fieldcorrHH = 4 \Theta(T) \Cabsmmconj 
\end{soe}.
\end{equation}
with $\mat{A}^*$ the element-wise complex conjugate of $\mat{A}$. 

Equation \eqref{eq:10} means that the complex correlation of the fields emitted by the structure is related to the complex coupling between impressed electric and magnetic current sources when computing the power absorbed by the structure when it is excited by these currents.

Once the correlation function of the fields is obtained, one can apply Wolf's modal decomposition of the fields \cite{16} to retrieve the fully coherent modes through which the structure is emitting fields and deal with them separately since they are adding up incoherently.

\section{Numerical computation of emissive modes}
To compute the modal content of the fields emitted by a lossy structure, one needs to compute the power absorbed by the structure when it is excited by electric or magnetic currents. The Method of Moments (MoM) is well-suited for such study since it routinely deals with both types of currents \cite{17}.

\subsection{The Method of Moments}
The MoM is a frequency domain integral equation-based method. It exploits the equivalence principle, which tells us that the fields scattered by a dielectric contained within a volume can be modeled using equivalent currents on the surface of this volume. Thus, instead of computing directly the fields scattered by a homogeneous piece of material, one can reformulate the problem to find the equivalent currents on the surface of that material. These unknown equivalent currents can be found by imposing the proper boundary conditions across the surface of the volume. One popular set of boundary conditions that is used consists in the continuity of the tangential electric and magnetic fields, leading to the so-called PMCHWT formulation \cite{17}. A key point of the MoM is that, since the equivalence principle is used, the problem can be divided into two subproblems. In the outer subproblem, the piece of material is removed and modeled using equivalent currents on its surface. In the inner problem, the whole space is filled with the given material and the ``free-space" is modeled using opposite equivalent currents on the same surface. In this way, in each subproblem, the corresponding homogeneous Green's functions can be used to compute the fields radiated by the equivalent currents \cite{17}.

In order to implement the MoM, the surface of the scatterer is first discretized using basis and testing functions. The unknown electric and magnetic currents are expanded using the finite set of basis functions while the boundary conditions are imposed in a weak sense using the finite set of testing functions. 

Considering the set $\{\vect{f}_B^i(\rpos)\}$ made of $2N$ basis functions, the unknown equivalent electric and magnetic currents read
\begin{equation}
\begin{soe}
\vect{J}_{eq}(\rpos) \simeq \sum_{i=1}^N x_i \vect{f}_B^i(\rpos),\\
~ \vspace{-0.3cm}\\
\vect{M}_{eq}(\rpos) \simeq \sum_{i=N+1}^{2N} x_i \vect{f}_B^i(\rpos).
\end{soe}
\end{equation} 
The unknown coefficients $x_i$ are determined by imposing the continuity of the tangential electric and magnetic fields across the interface, leading to the system of equations
\begin{equation}
\label{eq:11}
\mat{Z} \cdot \vect{x} = \vect{b}
\end{equation}
with $\vect{x}$ a vector containing the unknowns, $\vect{b}$ a vector related to the discontinuity of the total fields induced by the incident fields along the surface of the volume and $\mat{Z} = \mat{Z}_{in} + \mat{Z}_{out}$ with $\mat{Z}_{in}$ (resp. $\mat{Z}_{out}$) the impedance matrix whose $(i,j)$ entry corresponds to the fields projected on testing function $i$, as radiated by the currents on basis function $j$ for the inner (resp. outer) problem. Note that, as is often the case, the singular contribution of magnetic and electric basis functions on overlapping electric and magnetic testing function, respectively, has been excluded from the inner and outer impedance matrices since it is canceling out in \eqref{eq:11} \cite{17}.

As explained in \cite{18}, the power dissipated by a structure is equal to the power dissipated by the equivalent currents on the surface of the structure. Considering that all the sources are located outside of the emitter, one can compute the power dissipated by the latter using the formula
\begin{equation}
P_\text{tot} = -\dfrac{1}{4} \vect{x}^\dagger \cdot \Big(\mat{Z}^\text{in} + \mat{Z}^{\text{in}, \dagger}\Big) \cdot \vect{x}
\end{equation}

\subsection{Computation of the emissive modes}
In order to compute the emissive modes of the structure studied, we embedded it into a larger spherical surface, which was discretized to provide a discrete basis to excite the structure using impressed currents.

Denoting the surfaces of the structure and of the exciting sphere with the subscripts $1$ and $2$, respectively, and noting $\mat{Z}_{ij}$ the impedance matrix providing the fields induced in surface $i$ by currents on surface $j$, one can compute the discretized version $\Cabsnor$ of the correlation matrix of \eqref{eq:12} as
\begin{equation}
\label{eq:13}
\Cabsnor = -\dfrac{1}{4} ~ \Zij{12}^\dagger \cdot \big(\Zij{11}^{-1}\big)^\dagger \cdot \big(\Zijin{11} + \Zijin{11}^\dagger\big) \cdot \Zij{11}^{-1} \cdot \Zij{12}.
\end{equation}

Then applying the reciprocity relation of \eqref{eq:10}, one can compute the discrete version of the correlation functions of the fields as
\begin{equation}
\label{eq:14}
\fieldcorrnor = 4 \Theta(T) ~ \mat{S} \cdot \Cabsnor^* \cdot \mat{S}
\end{equation}
with $\mat{S}$ a diagonal matrix whose diagonal entries are $\pm1$ and that is used to switch the sign of the entries of the $\Cabsnor^{JM}$ and $\Cabsnor^{MJ}$ blocks \cite{18}.

Finally, Wolf's modal decomposition of the fields can be applied by computing the eigenvectors and corresponding eigenvalues of $\fieldcorrnor$. The eigenvectors $\vect{v}_i$ correspond to the fully coherent field distribution of each mode tested on the basis functions of the exciting spheres, while the corresponding eigenvalues $e_i$ provide the squared amplitude of the mode. 


\subsection{Computation of the radiated power}
The modes adding up incoherently, the powers transported by each mode can be computed independently and then summed up, the coupling between different modes vanishing. Using the exciting sphere as a fictitious interface where equivalent currents can be computed, on can find that the power transported by mode $i$ is given by

\begin{equation}
\label{eq:15}
P_i = -\dfrac{e_i}{4} \vect{v}_i^\dagger \cdot \big(\Zij{22}^{-1}\big)^\dagger \cdot \big(\Zijout{22} + \Zijout{22}^\dagger\big) \cdot \Zij{22}^{-1} \cdot \vect{v}_i.
\end{equation}

Noting that $\Zijin{22} = \Zijout{22}$ since both sides of the exciting sphere are corresponding to free-space, \eqref{eq:15} can be simplified to give
\begin{equation}
P_i = -\dfrac{e_i}{8} \vect{v}_i^\dagger \cdot \Big(\big(\Zij{22}^{-1}\big)^\dagger + \Zij{22}^{-1}\Big) \cdot \vect{v}_i.
\end{equation} 

\section{Numerical validation}
\begin{figure}
\center
\includegraphics[width = 9cm]{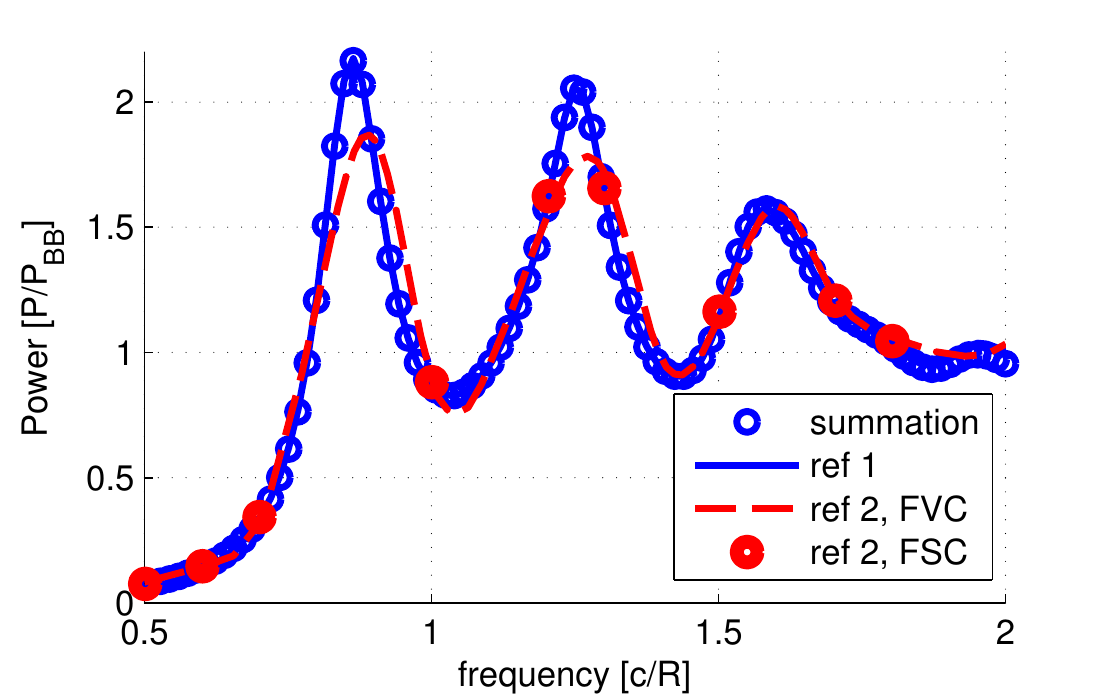} \\
(a) \\
\includegraphics[width = 9cm]{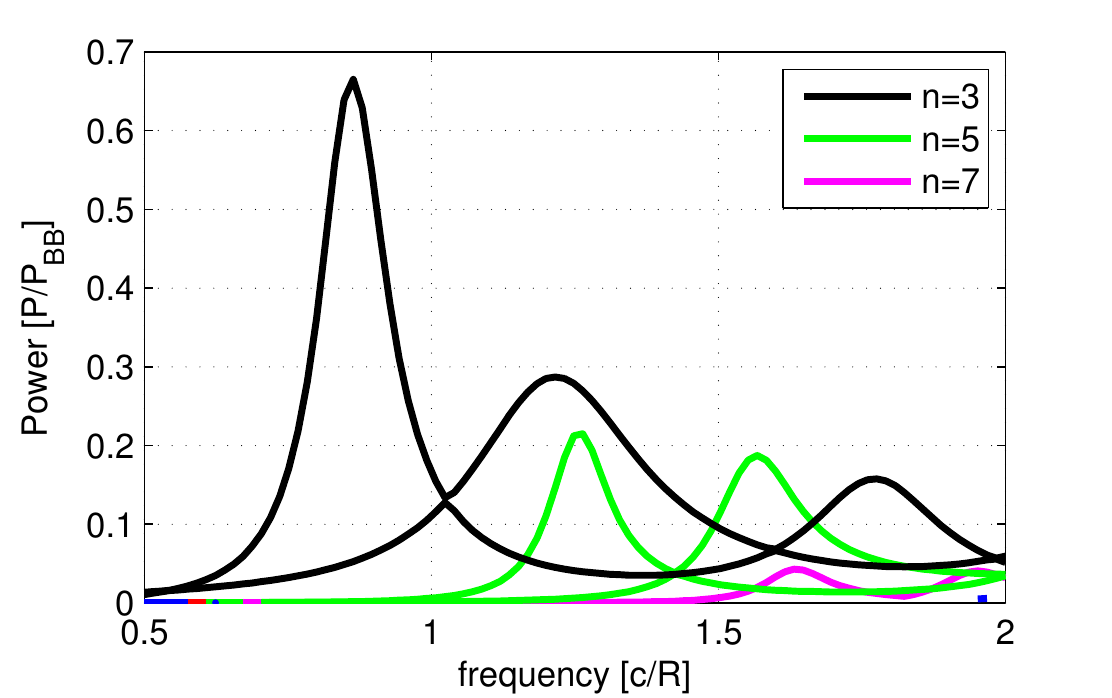} \\
(b)
\caption{Normalized power emitted by the dielectric lossy sphere. The power has been normalized by $P_{BB} = \dfrac{A}{4\pi}(\omega/c)^2 \Theta(T)$, with $A$ the area of the surface of the sphere, in order to ease the comparison with \cite{7}. (a) Total power emitted. (b) modal distribution of the power emitted. The colors indicate the degeneracy $n$ of the mode.}
\label{fig:01}
\end{figure}

\begin{figure}
\center
\includegraphics[width = 9cm]{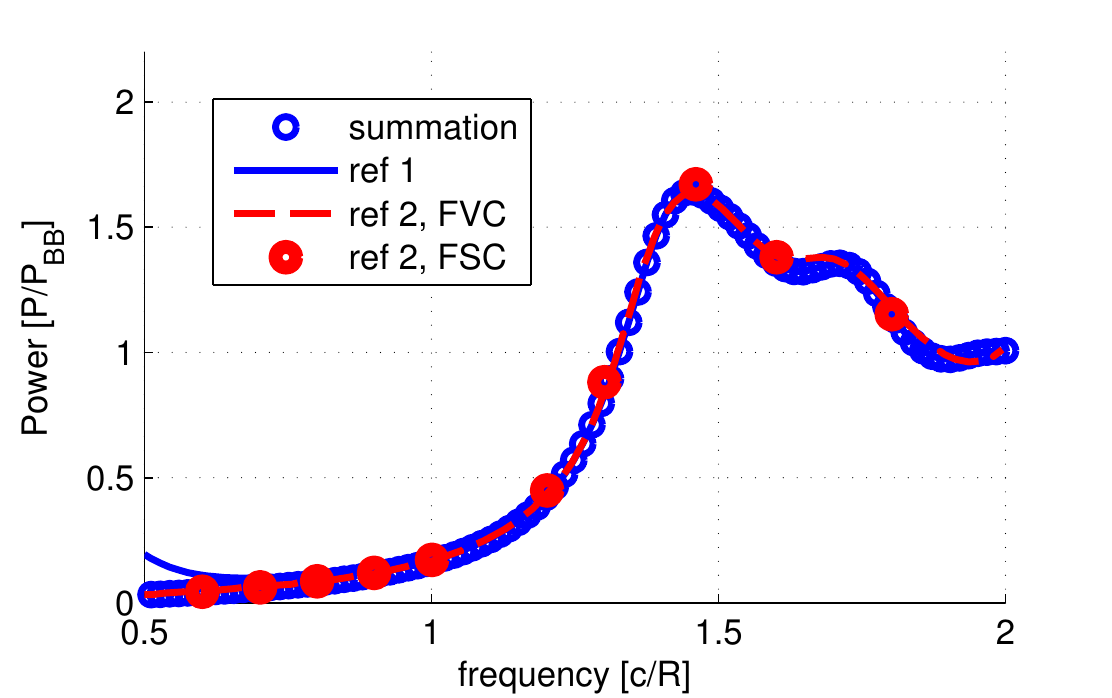} \\
(a) \\
\includegraphics[width = 9cm]{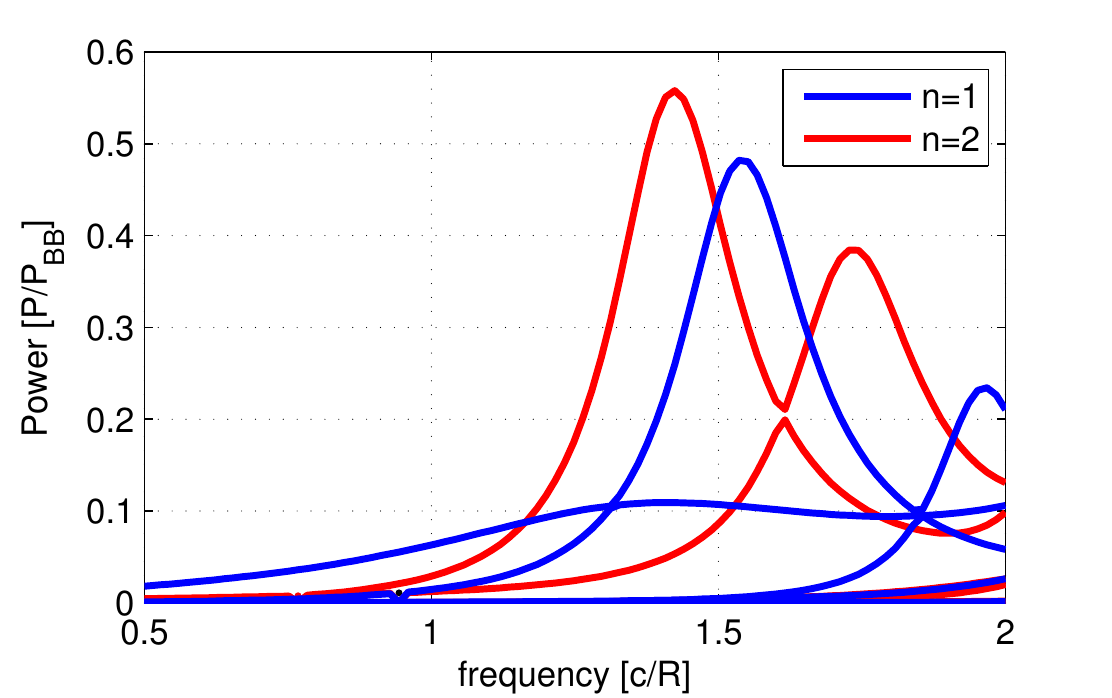} \\
(b)
\caption{Normalized power emitted by the dielectric lossy ellipsoid. The power has been normalized by $P_{BB} = \dfrac{A}{4\pi}(\omega/c)^2 \Theta(T)$, with $A$ the area of the surface of the sphere, in order to ease the comparison with \cite{7}. (a) Total power emitted. (b) modal distribution of the power emitted. The colors indicate the degeneracy $n$ of the mode.}
\label{fig:02}
\end{figure}

To implement the formalism proposed in the previous Section, we used a MoM in-house code. We computed the power radiated by a lossy sphere and a lossy ellipsoid of relative permittivity $\varepsilon_r = 12-1j$ as a function of the frequency, as done in \cite{7}. Note that the imaginary part of the permittivity can be related to a real conductivity through the relation $\sigma = -\omega \varepsilon_0 \Im(\varepsilon_r)$, with $\Im(a + jb) = b$, $a$ and $b$ being real numbers, and $\varepsilon_0$ the free-space permittivity. The radius of the sphere is $R=1$ (arbitrary unit). The ellipsoid is obtained by shrinking the sphere by a factor of two along two orthogonal directions.
The sphere and the ellipsoid are meshed using 2118 and 1032 RWG basis functions \cite{20}, respectively. The fictitious sphere used to excite the structure has a radius of 2 and is meshed with 2118 RWG basis functions.
The total power radiated by the sphere and the ellipsoid are visible in Figs. \ref{fig:01}(a) and \ref{fig:02}(a), respectively, and their corresponding eigenmode decomposition are visible in Figs. \ref{fig:01}(b) and \ref{fig:02}(b), respectively.

In order to validate the presented formalism, we used three different techniques to compute the total power emitted by the structure. First, we used formula (21) of \cite{6} (ref 1). Second, we compared with the results presented in \cite{7}, which they computed using both surface (ref2, FSC) and volume (ref2, FVC) integral equations. Last, we used the modal decomposition of the fields described in the previous Section and summed the contribution of each mode (summation). 

{ 
It can be seen that the correspondence between the formula of \cite{6} and the modal decomposition is very good. A small deviation appears at low frequencies for the ellipsoid, probably originating from the low-frequency instability of the MoM. This instability does not appear in the modal decomposition since fields are first propagated to the fictitious sphere before being decomposed into modes. This extra propagation step tends to regularize the results by filtering out part of the reactive field. }

{ 
Comparing with the results of \cite{7}, we obtain a very good agreement with the results that they obtained using surface integral equations. However, some differences with the volume integral equation formulation are visible near resonance. These are probably originating from numerical method used rather than the formulation itself.}

Concerning the modal content of the fields emitted by the sphere (Fig. \ref{fig:01}(a)), it can be seen that for wavelengths that are large with respect to the size of the emitter, the power is mainly emitted through three degenerate modes. For increasing frequencies, the electrical size of the sphere increases so that the number of significant modes increases too. However, higher-order resonances of the sphere are clearly distinguishable. Thus, even for high frequencies, the total power emitted by the sphere may mainly come from the contribution of few degenerated resonant modes. It can also be noticed that the degeneracy of the modes ($n=3,5,7$) is typical for a spherical geometry admitting spherical harmonics as solutions.

Looking at the modal content of the fields emitted by the ellipsoid (Fig. \ref{fig:02}(b)), we can see that the lowered symmetry gives rise to decreased degeneracy of the modes, as one can expect. Moreover, the first resonant mode appears for much shorter wavelength due to the smaller size of the structure.

These results seem to indicate that, controlling the resonant frequencies of the structure, the number of emitting modes can be tailored.

\section{Conclusion}
In this paper, we proposed to characterize thermal emitters using a modal decomposition of their emitted fields. The partially coherent fields are corresponding to an incoherent superposition of this fully coherent modes. To compute the emissive modes of a structure, the correlation function of the emitted fields is computed using reciprocity. Then, the emissive modes are obtained by applying an eigenmode decomposition of the correlation function.
This modal decomposition has been implemented using a MoM software and tested on a sphere and an ellipsoid. It is shown that the number of emissive modes depends on the electrical size of the emitter. This indicates that, by properly designing the structure, the number of emissive modes at a given frequency can be tailored.


\section*{Acknowledgment}
This project has received funding from the European Union's Horizon 2020 research and innovation programme under the Marie Sk\'{l}odowska-Curie grant agreement No 842184.

{ 
The authors would like to thank prof. J.-J. Greffet for the interesting discussions they had and for pointing out ref. \cite{Rytov}.
}



%

\end{document}